# Real-space tilting method for atomic resolution STEM imaging of nanocrystalline materials

Jiake Wei[1]*, Zhangze Xu[1,2], Wenjie Shen[1,2], Bin Feng[3,4], Ryo Ishikawa[3], Naoya Shibata[3], Yuichi Ikuhara[3], Xuedong Bai[5]*

[1] State Key Laboratory of Catalysis, Dalian Institute of Chemical Physics, Chinese Academy of Sciences, Dalian 116023, China.

[2] School of Chemistry, Dalian University of Technology, Dalian 116024, China.

[3] Institute of Engineering Innovation, The University of Tokyo, Tokyo 113-8656, Japan.

[4] PRESTO, Japan Science and Technology Agency, Kawaguchi, Saitama 332-0012, Japan.

[5] State Key Laboratory for Surface Physics, Institute of Physics, Chinese Academy of Sciences, Beijing, 100190, China.

*Correspondence to: J. W. (weijiake@dicp.ac.cn), X. B. (xdbai@iphy.ac.cn)

# Abstract

Atomic-resolution scanning transmission electron microscopy (STEM) characterization requires precise tilting of the specimen to high symmetric zone axis, which is usually processed in reciprocal space by following the diffraction patterns. However, for small-sized nanocrystalline materials, their diffraction patterns are too faint to guide the tilting process. Here, a simple and effective tilting method is developed based on the diffraction contrast change of the shadow image in the Ronchigram. We can calculate the misorientation angle of the specimen and tilt it to the zone axis based on the position of the shadow image with lowest intensity. This method requires no prior knowledge of the sample and the maximum misorientation angle we can correct is greater than ±6.9° with sub-mrad accuracy. It is processed in real space, without recording the diffraction patterns of the specimens, which can effectively apply to nanocrystalline materials. Combined with the scripting to control the microscope, we can automatically tilt the sample to the zone axis under low dose condition (<0.17 e$^-$/Å$^2$/s), which could facilitate the imaging of beam sensitive materials such as zeolites or metal organic frameworks. This automated tilting method could contribute to the atomic-scale characterization of the nanocrystalline materials by STEM imaging.

# Introduction

State-of-the-art scanning transmission electron microscopy (STEM) is one of the most powerful atomic resolution characterization tools in materials science reserach[1-3]. By employing an atomically focused electron beam to scan across the specimen, STEM experiments now enable us to visualize individual atoms and their arrangements within materials[3-10]. Combined with electron energy loss spectroscopy and energy dispersive X-ray spectroscopy, STEM could characterize their chemical compositions, electronic structures and even phonon structures at atomic resolution[11-14]. STEM is a projected imaging technique, and the atomic resolution analyses require precisely tilting the specimen into a high symmetric zone axis. This alignment ensures that atoms are well-ordered along the optical axis, facilitating column-by-column imaging and spectroscopy.

Tilting the sample to the zone axis in STEM is a time-consuming aspect, if not impossible, for nanocrystalline materials. Conventionally, tilting the specimen to the zone axis is processed in reciprocal-space by following the selected area electron diffraction (SAED) in parallel TEM mode or Kikuchi pattern under convergent-beam electron diffraction (CBED) in the Ronchigram[15-17]. However, these methods face limitations when applied to nanocrystalline materials, particularly for those smaller than 40-50 nm. This is because the diffraction patterns of individual nanocrystalline materials are typically too faint to guide the tilting process (See an example in Figure S1). Meanwhile, to obtain the Kikuchi pattern usually requires higher beam currents, which may have the risk of damaging the nanomaterials, since they could be fragile under electron beam irradiation[18,19]. Furthermore, the Kikuchi patterns may not even manifest because they can only form from specimens with enough thickness[2]. Until today, to image materials less than 40 nm, we usually try to randomly find a near-zone axis specimen from large number of samples, which wastes time and resources.

In this study, we have developed a simple and effective real-space tilting method for nanocrystalline materials based on the diffraction contrast change of the largely defocused shadow images in the Ronchigram[1]. When the incident electron beam is parallel to the high symmetric zone axis of the specimen, the incident electrons would be mostly scattered because of satisfying the Bragg diffraction condition, and hence, their contrast will change in the shadow image (formed by the directly transmitted electrons) compared to those acquired under off-zone axis conditions[2,20]. We show that we can calculate the misorientation angle of the specimen and tilt it to the zone axis based on the position of its darkest shadow image in the Ronchigram. This tilting method has no requirements of prior knowledge of the specimen and the maximum misorientation angle we can correct could be greater than ±6.9° with an accuracy of sub-mrad. It is processed in real space without recording the diffraction patterns of the samples, thus can effectively apply to nanocrystalline materials. By combining with the scripting to control the microscope, we can automatically tilt the sample to the zone axis under low dose condition (~0.17 $e^-/Å^2/s$), which could potentially facilitate the imaging of beam sensitive materials.

## Results

Figure 1a depicts a simplified electron ray-path diagram of the Ronchigram, which is widely employed for correcting the residual aberrations and finding the target specimens before STEM observation[1]. In Ronchigram, a shadow image of the specimen is formed at the detector plane by placing the sample away from the focal point. Figures 1d-f show the shadow images of a $γ-Al_2O_3$ nanorod, marked by the dotted green lines. The rod is oriented along the [100] zone axis of $γ-Al_2O_3$, which is confirmed by the annular dark field (ADF)-STEM images, provided in Figures 1b and 1c. When the rod is in-plane moved to different positions relative to the optical center (the center of the red circles in Figures 1d-f), the contrast of the nanorod varies from Figures 1d-f. Figure 1g presents the intensity line profiles of the shadow images in Figures 1d-f of the rod (from "A" to "B"). The line profiles distinctly reveal that the image in Figure 1e, where the target specimen aligns well along the optical center, exhibits lower intensity compared to Figures 1d and 1f, where the specimen/image is displaced from the optical center.

The contrast variation observed in the shadow images of Figures 1d-f can be interpreted through the diffraction contrast. As shown in Figure 1a, when the specimen is on the optical center, the incident electrons, predominantly following the vertical dotted blue line, is parallel to the zone axis, thus satisfying the Bragg condition. Consequently, significant electron scattering occurs, leading to a decrease of the number of the directly transmitted electrons, which shows lower intensity of the shadow image as demonstrated in Figure 1e. Conversely, when the specimen location deviates from the optical center (Figures 1d and 1f), the incident electrons are illuminated mainly along the pink dotted lines in Figure 1a. Under this scenario, the incident electron beam is no longer parallel to the zone axis, resulting in a higher intensity in the shadow image compared to Figure 1e. Based on the above analysis, it can be inferred that the zone axis specimen should have the darkest shadow image (with lowest intensity) when the specimen is placed on the optical center.

On the contrary, if we encounter a specimen exhibiting the darkest image with lowest intensity at a location other than the optical center in the Ronchigram as exemplified in Figure 2a, we can derive the misorientation angle from the zone axis. For instance, if the darkest shadow image appears to the right-hand of the optical center, as depicted in Figure 2a, it indicates when the incident beam propagates along the dotted pink line, it aligns parallelly to the zone axis of the sample. Consequently, we can calculate the misorientation angle of the specimen using the equation:

$$\theta = \arctan(d/c), \qquad (1)$$

where $d$ is the distance between the darkest shadow image and the optical center, and $c$ denotes the camera length, which is the distance between the in-focused sample plane (focal point) to the detector plane. Subsequently, tilting the sample by $\theta$ (along the β rotation axis as shown in Figure 3) will bring the darkest shadow image to the optical center shown in Fig. 2b, suggesting this specimen is tilted to the zone-axis. From an experimental standpoint, it is more practical to calibrate the shadow image in an angular unit (e.g., mrad) by employing the CBED pattern of a known specimen.

An experimental demonstration of the tilting is presented in Figures 2c and 2d, where we tried to tilt the nanorod in the dotted green lines to the zone axis. Initially, we sought to locate the position of the darkest shadow image of the nanorod in the

Ronchigram by in-plane moving the specimen stage (Figure S2). The contrasts of the shadow image varied with the different image positions, and the darkest image was identified in Figure 2c (Figure S2d). The darkest position of the image was estimated to be 63.6 mrad away from the optical center. Subsequently, after tilting the sample by 63.6 mrad, the darkest shadow image coincided with the optical center in Figure 2d (Figure S3). This alignment indicates that the sample is under the zone axis condition.

It is worth noting that tilting the specimen to the zone axis typically necessitates a double tilting holder, which incorporates two tilting axes, as depicted in Figure 3a. Conventionally, the tilting axes of α and β are aligned parallel to the x and y axes of the holder, respectively. As illustrated in Figures 2a-b and 3a, when moving the specimen (or its shadow image) along the x direction, it is equivalent to tilting the sample along the β rotation axis. Similarly, tilting the sample along the α axis corresponds to shifting the sample along the y direction. In other words, tilting the sample along the α and β rotation axes leads to the displacement of the darkest shadow image along the y and x directions in the Ronchigram, respectively. This underlying principle streamlines the tilting process, as demonstrated in Figure 3b-c. Through in-plane moving of the specimen stage, it was found that the sample (γ-$Al_2O_3$ nanorod) in the blue dotted circle has the darkest shadow image on the position in Figure 3b. The vector from the darkest image position to the optical center can be decomposed into two components (illustrated by pink arrows in Fig. 3b) along the x and y directions, respectively. Tilting the sample by 88.6 mrad and -14.6 mrad in the α and β rotation axes, respectively, would orient it to the zone axis, as depicted in Figures 3c and 3e-f. The ADF-STEM image captured before tilting (Figure 3d) only shows the shape of the rod without any atomic-scale information. On the other hand, the images acquired after tilting in Figures 3e and 3f clearly indicate that the nanorod is aligned on the [011] zone axis and the atomic arrangements of rod, including the bulk and surface atomic structures, were revealed.

It is important to note that the tilting process can also be automatically performed through the scripting in available commercial software (e.g. the Digital Micrograph software[21]) to control the microscope, as demonstrated in Figure 4 and Movie S1. As shown in Movie S1 and Figure 4a-c, to automatically tilt the specimen to the zone axis, it is firstly needed to find a target sample (e.g. a ZSM-5 nanosheet in the blue circle of Figure 4a) and locate its darkest shadow image (the center of the blue circle in Figure 4b) by moving the specimen stage. Once finding the darkest image, it is possible to automatically calculate the misorientation angle and control the holder to tilt the specimen to the zone axis as shown in Figure 4c and Movie S1. The ADF-STEM images (Figures 4d and 4f) acquired immediately after the auto-tilting process and the corresponding FFT pattern (Figure 4e) suggest the nanosheet was aligned on the [010] zone axis of ZSM-5. The whole tilting process can be completed within 20 seconds.

This auto-tilting process could facilitate the imaging of beam-sensitive materials such as zeolites. Since the tilting method relies on contrast changes of largely defocused shadow image, the electron dose can be kept low compared to conventional tilting methods in STEM, like following the Kikuchi pattern in CBED, where the magnifications must be high enough. For instance, in Figures 4a-c and Movie S1, the defocused value was set to be ~20 μm, and the probe current was measured to be ~146 pA. Therefore, the electron dose rate for tilting is estimated to be 0.5 $e^-/Å^2/s$, resulting in a dose of ~10 $e^-/Å^2$ for the whole tilting process. This dose can be neglected compared to the one required to image the sample at atomic resolution in Figure 4f (~3000 $e^-/Å^2$). Moreover, the electron dose rate for tilting can further be reduced (e.g. ~0.17 $e^-/Å^2/s$ as shown in Figure S4) by using larger defocus value or lower probe current.

**Discussion**

The accuracy of this tilting method relies on the precise identification of the position of the darkest shadow image, which is influenced by various factors such as the defocused value (magnification of the shadow image), probe current (signal-to-noise ratio of the image), and the symmetry of the zone (rate of the image intensity changes with increasing misorientation angle). As schematically shown in Figure 5a, to image the different positions of a specimen, the incident electrons are propagated with different misorientation angle from the optical axis, which results an intensity change of the shadow image within the sample as shown in Figures 5b and 5c. Figure 5f givens the intensity line profile along the yellow arrow in Figure 5c, where the area of the nanosheet is yellow shadowed. Clearly, the intensity is increasing when the misorientation angles increase. The line profile can be fitted by a Gassian function (the red curve in Figure 5f), and the standard error of the position with the minimum intensity is estimated to be 0.03 mrad, indicating that a theoretical accuracy of ~0.002° can be achieved when determining the darkest position. However, in practice, the tilting accuracy could be influenced by the quality of the shadow image and the precision of the observer's judgment by naked eyes. Nevertheless, based on our experimental experiences, an accuracy of <0.5 mrad can be usually obtained. To achieve better tilting accuracy, it is advisable to position the sample closer to the focal point (higher magnification) and use a higher probe current to obtain shadow images with higher signal-to-noise ratio (See an example in Figure S5). Additionally, the maximum misorientation angle of the above tilting method is dependent on the convergence angle of the illumination beam. For instance, with a convergence semi-angle of 120 mrad, the maximum misorientation angle could be ±6.9°. This implies that if the sample is within ±6.9° from the zone axis, its darkest shadow image position can always be found within the field of view of the Ronchigram, enabling it to be tilted to the zone axis. The maximum misorientation angle could be further enlarged by using a larger convergence angle (larger condenser lens aperture).

In summary, we have developed a straightforward real space tilting method based on the diffraction contrast change of shadow images in the Ronchigram. This method does not require any prior knowledges of the specimen and allows for a maximum misorientation angle of ±6.9°. By integrating scripting to control the microscope, it becomes feasible to automatically tilt the sample to the zone axis. The tilting approach here relies on contrast change in the shadow image and does not necessitate the recording of the SAED or CBED patterns of the specimen. Consequently, it can be applied to materials with small sizes (e.g., <40 nm), where their single crystal SAED and CBED are too weak to guide the tilting process. Furthermore, this tilting method is processed under the Ronchigram mode in STEM, obviating the need for frequent switching between STEM and TEM modes as required by conventional tilting method following the SAED. Moreover, compared to tilting methods following the Kikuchi pattern in CBED, our method requires lower electron doses, making it suitable for tilting beam-sensitive materials such as zeolites and metal organic frameworks. Overall, our tilting method offers a time- and resource-saving approach for imaging nanocrystalline materials such as semiconductors[22], catalysts[4,23,24], and battery materials[25]. It should help the atomic scale understanding of material structures, particularly for nanocrystalline materials. In addition, the automatically tilting method proposed could also contribute to building the next generation AI driven automating microscope[26].

# Figures and figure legends

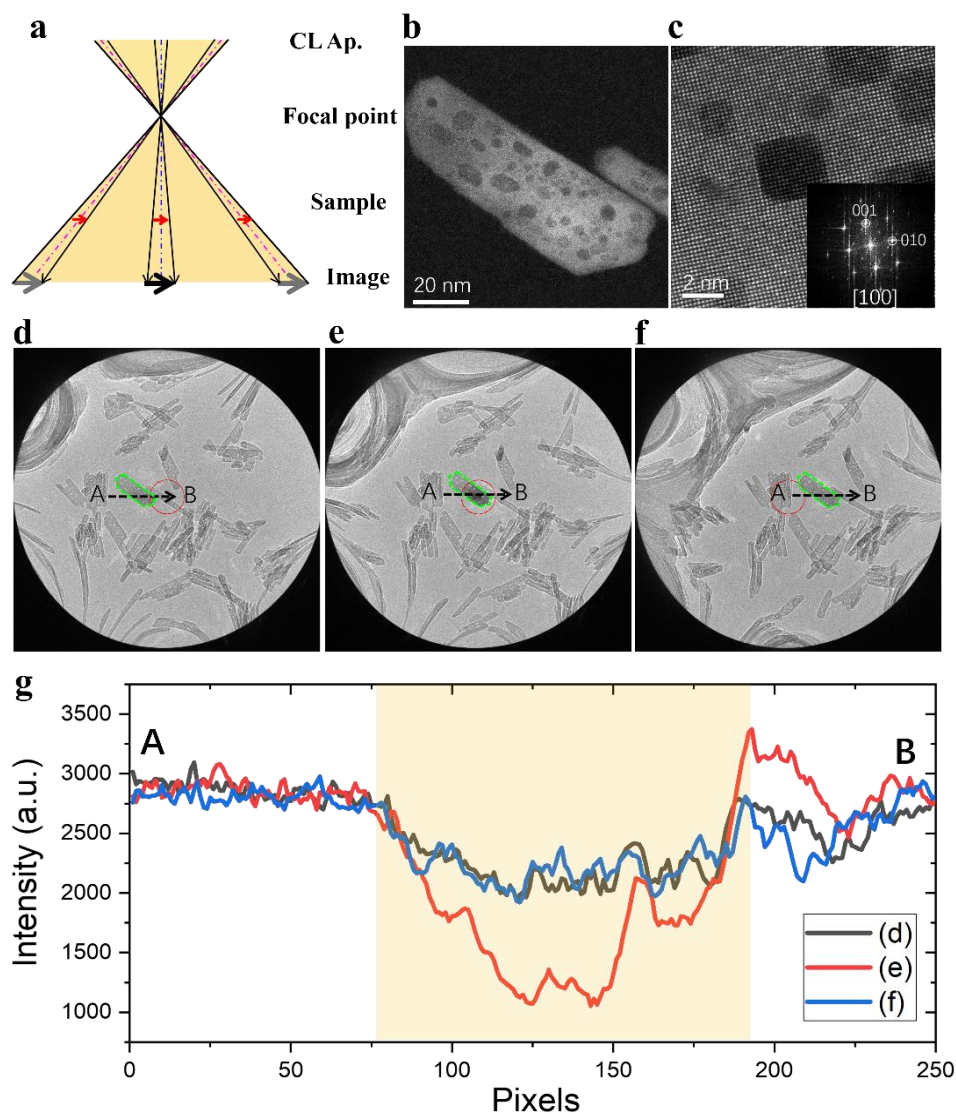

**Figure 1. Diffraction contrast variation of the largely defocused shadow image in Ronchigram.** (a) The electron ray-path diagram of the formation of shadow image in Ronchigram with a sample on the zone axis. The red arrows indicate the specimen, positioned beneath the focal point of the condenser lens, while the black and gray arrows represent the shadow images of the specimen. **(b-c)** The ADF-STEM images of the γ-$Al_2O_3$ nanorod, marked by green lines in (d-f). The fast Fourier transform (FFT) of (c) is presented in the inset of the image. The nanorod is on [100] zone axis. **(d-f)** The shadow images of the nanorod at different positions, with the defocus set to 3.5 μm. **(g)** The intensity line profile (from A to B in (d)) of the shadow images in (d-f). The yellow shadowed area represents the position of the nanorod.

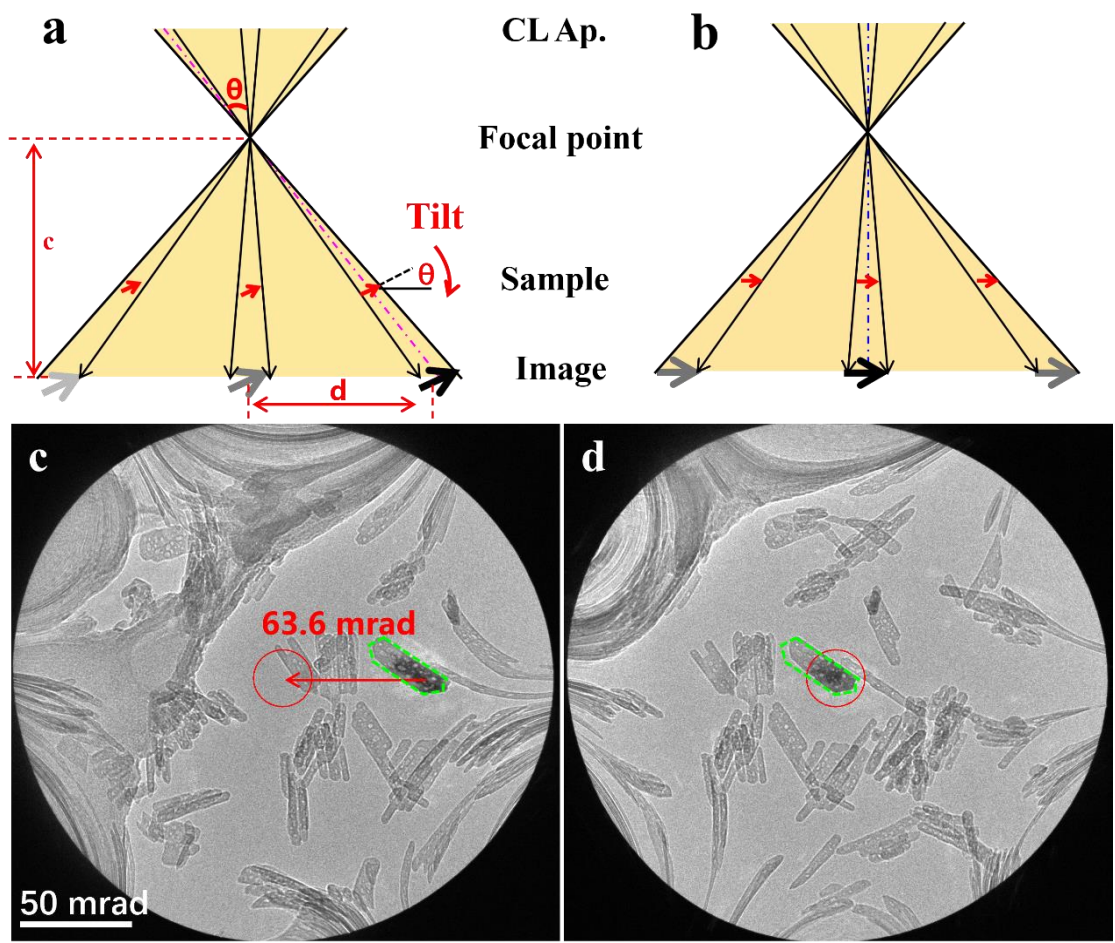

**Figure 2. Demonstration of the tilting method based on the contrast change of the shadow images.** (**a**) The electron ray diagram of the shadow image with an off-zone axis specimen. The darkest shadow image is not located on the optical center. (**b**) The electron ray diagram of the shadow image of the specimen in (a) after tilting by θ. The darkest shadow image is coincident with the optical center. (**c**) Experimental darkest shadow image of the nanorod before tilting. The searching for this darkest image is given in Figure S2. (**d**) Experimental darkest shadow image of the nanorod after tilting. The searching for this darkest image is given in Figure S3.

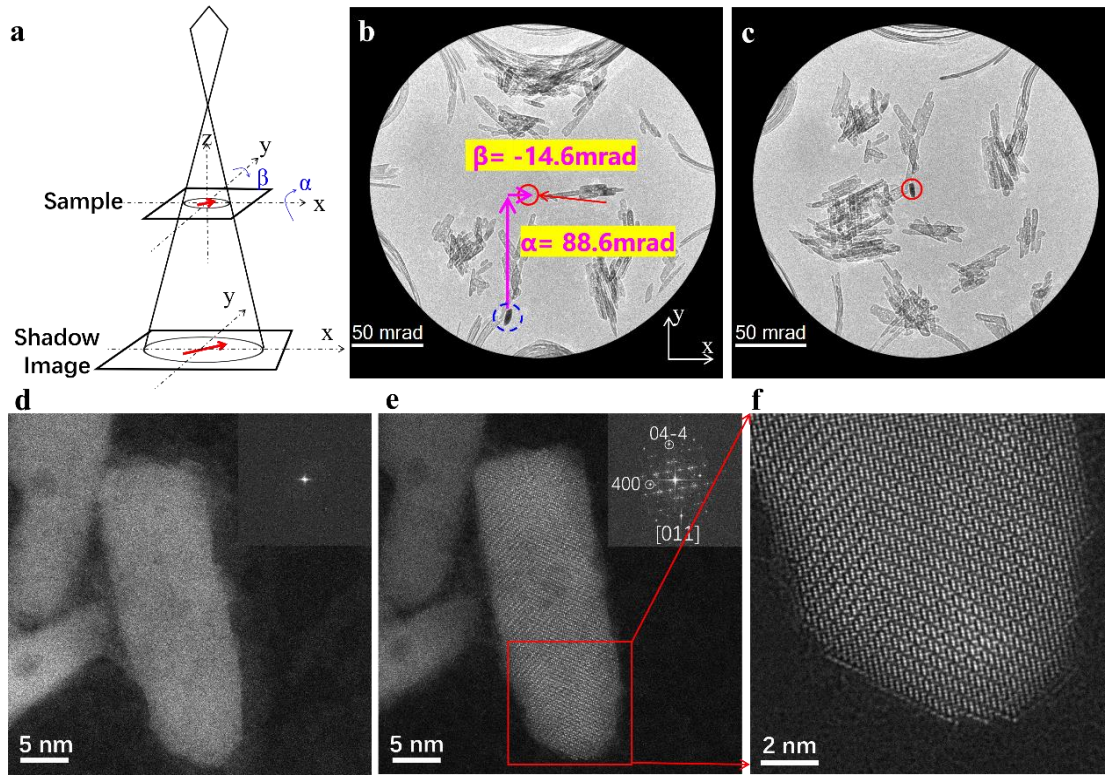

**Figure 3. Tilting the specimen to zone axis on a double-tilting holder.** (**a**) A schematical drawing showing the structure of a double-tilting holder. The tiling axes of α and β are aligned parallel to the x and y axes of the holder, respectively. (**b**) Experimental darkest shadow image of the γ-Al$_2$O$_3$ nanorod (marked by dotted blue circle) before tilting. The vectors from the darkest image to the optical center (the center of the red circle) could be decomposed to two vectors along the x and y direction with length of -14.6 mrad and 88.6 mrad, respectively. Rotating the specimen by 88.6 mrad and -14.6 mrad along the α and β, respectively, could tilt it to zone axis in (c). (**c**) The darkest shadow image of the nanorod after tilting. (**d-f**) The ADF-STEM images of the nanorod before (d) and after tiling (e and f). The FFT patterns of (d) and (e) were presented in the inset of the images. After tilting, the nanorod was aligned to the [011] zone axis of γ-Al$_2$O$_3$.

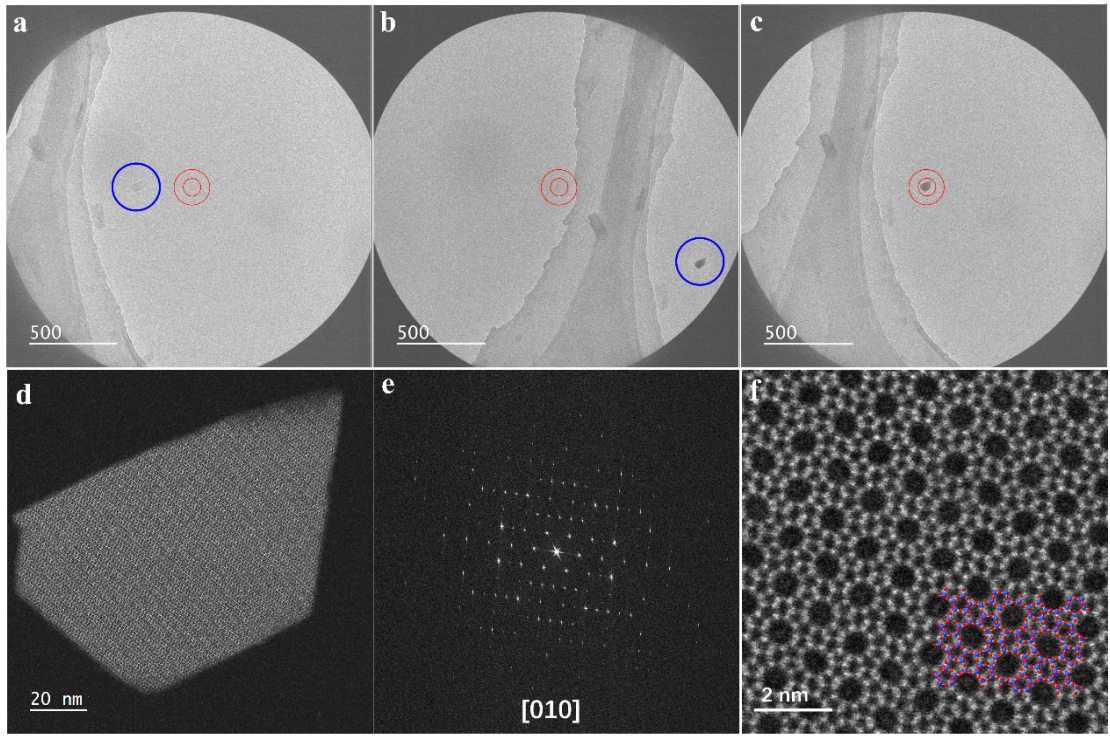

**Figure 4. Experimental demonstration of automatically tilting the ZSM-5 zeolites.** (**a-c**) Serial shadow images taken from Movie S1, which show the automatically tilting of the ZSM-5 nanosheet in the center of the blue circle in (a). The defocus value was set to be ~ 20 μm. The center of the red circles is the optical center. (**d**) The ADF-STEM images of the ZSM-5 nanosheet in (a-c) acquired immediately after the auto tilting shown in Movie S1. The sample was tilted to [010] zone axis, with the FFT pattern and magnified image given in (e) and (f), respectively. (**e**) The FFT pattern of the ADF image in (d). (**f**) The atomic resolution ADF-STEM image of the ZSM-5 nanosheet. The projected atomic model of ZSM-5 was overlaid on (f), where the blue and red spheres represent the Si and O atomic columns, respectively.

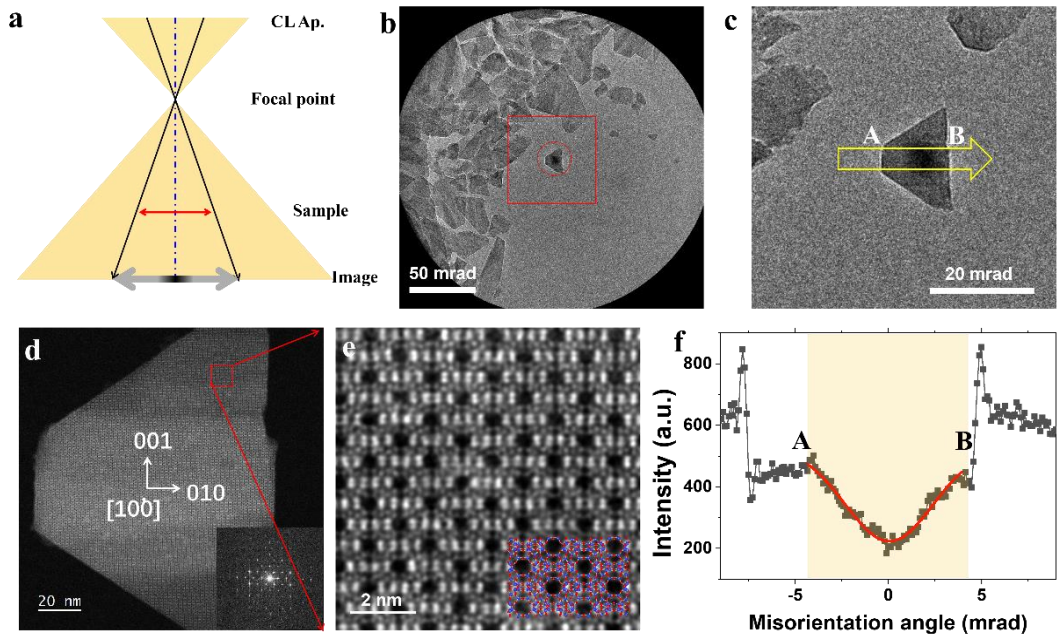

**Figure 5. The accuracy of locating the darkest position of the shadow image.** (**a**) The schematical drawing of the formation of the shadow image with contrast change within a sample. The red and black lines represent the specimen and its shadow image. For different position of the specimen, the imaged electrons were propagated with different misorientation, resulting in contrast variation of its shadow image. (**b, c**) Experimental shadow images of a ZSM-5 nanosheet. (c) is the magnified image from the red box in b. (**d, e**) The ADF-STEM images of the nanosheet in (b) and (c). (f)The intensity line profile along the white arrow in (c). The minimum intensity position is shifted to be 0 mrad. The line profile near the minimum intensity position is fitted by a Gaussian function (the red curve). The standard error of the minimum intensity position is estimated to be 0.03 mrad.

## Methods

**STEM characterization:** The synthesis of γ-Al$_2$O$_3$ nanorod and ZSM-5 nanosheet are follwing the privious reports[27,28]. The TEM sample was prepared by ultrasonically dispersing the specimen powder in ethanol, and droplets of the suspension were deposited on Cu grid with ultrathin carbon film. STEM images were recorded on a JEOL Grand ARM 300F microscope operating at 300kV. For acquiring the shadow images, a convergence semiangle of 120 mrad was employed. The shadow images were acquired by using the Gatan OneView camera.The ADF-STEM images were collected by setting the convergence semi-angle to be 16 mrad and the collection semi-angles are spanning from 27-110 mrad.

**Data availability:** The data that support the findings of this study are available within the article and its Supplementary Information. Any other relevant data are also available upon reasonable request from the corresponding authors.

**Code availability:** The code, which is used to automatically tilt the sample, is available from the corresponding author upon reasonable request.

**Funding:** This work was supported by the National Natural Science Foundation of China (12334001 and 51991344), National Key R&D Program of China (2021YFA1400204), and the Strategic Priority Research Program of Chinese Academy of Sciences (XDB33030200).

**Author contributions:** J. W. designed and carried out the experiment and wrote the manuscript. Z. X. synthesized the samples. X. B., W. S., B. F., R. I., N. S., and Y. I. discussed the results and commented on the manuscript.

**Competing interests:** The authors declare the following financial interests/personal relationships which may be considered as potential competing interests: J. W., Z. X. and W. S. are inventors on Chinese patent application (application No. 202311208642.6).

Supplementary Materials for

# Real-space tilting method for atomic resolution STEM imaging of nanocrystalline materials

Jiake Wei*, Zhangze Xu, Wenjie Shen, Bin Feng, Ryo Ishikawa, Naoya Shibata, Yuichi Ikuhara, Xuedong Bai*

*To whom correspondence should be addressed. J. W. (weijiake@dicp.ac.cn), X. B. (xdbai@iphy.ac.cn)

**This file includes:**

1. Supporting Figures and Figure Legends

2. Captions of Movies S1

**Other Supplementary Materials for this manuscript include the following:**

Movies S1

1. **Supporting Figures and Legends**

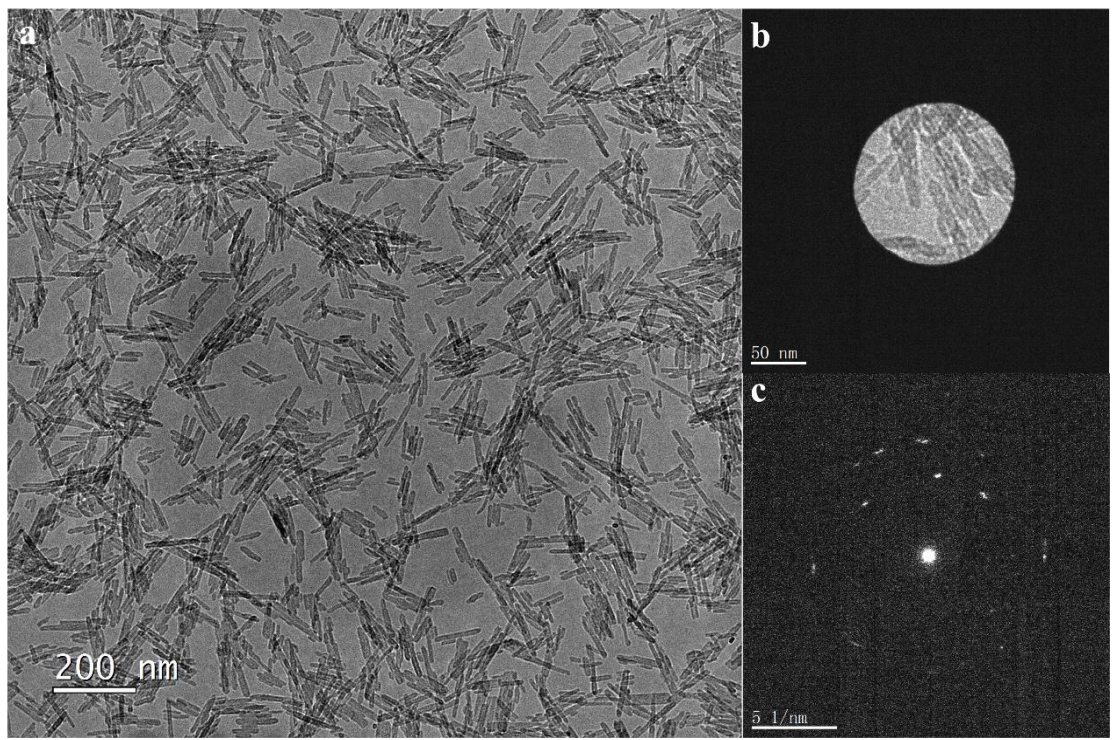

**Figure S1. An example of typical nanomaterial under transmission electron microscopy (TEM).** (**a**) The TEM image of the γ-alumina nanorods. (**b**) The smallest selected-area aperture available in the JEOL Grand ARM 300F instrument for selected area electron diffraction (SAED). (**c**) The SAED pattern obtained from the area in (b).

As shown in (a), the γ-alumina nanorods are closely adjacent to each other. It is challenged to acquire the diffraction pattern for individual nanorod even using the smallest selected-area aperture. Consequently, acquiring a single-crystal diffraction pattern becomes unfeasible. Even in instances where isolated nanorods are discernible, the resultant diffraction patterns often lack the intensity required for sample tilting, as demonstrated in (c).

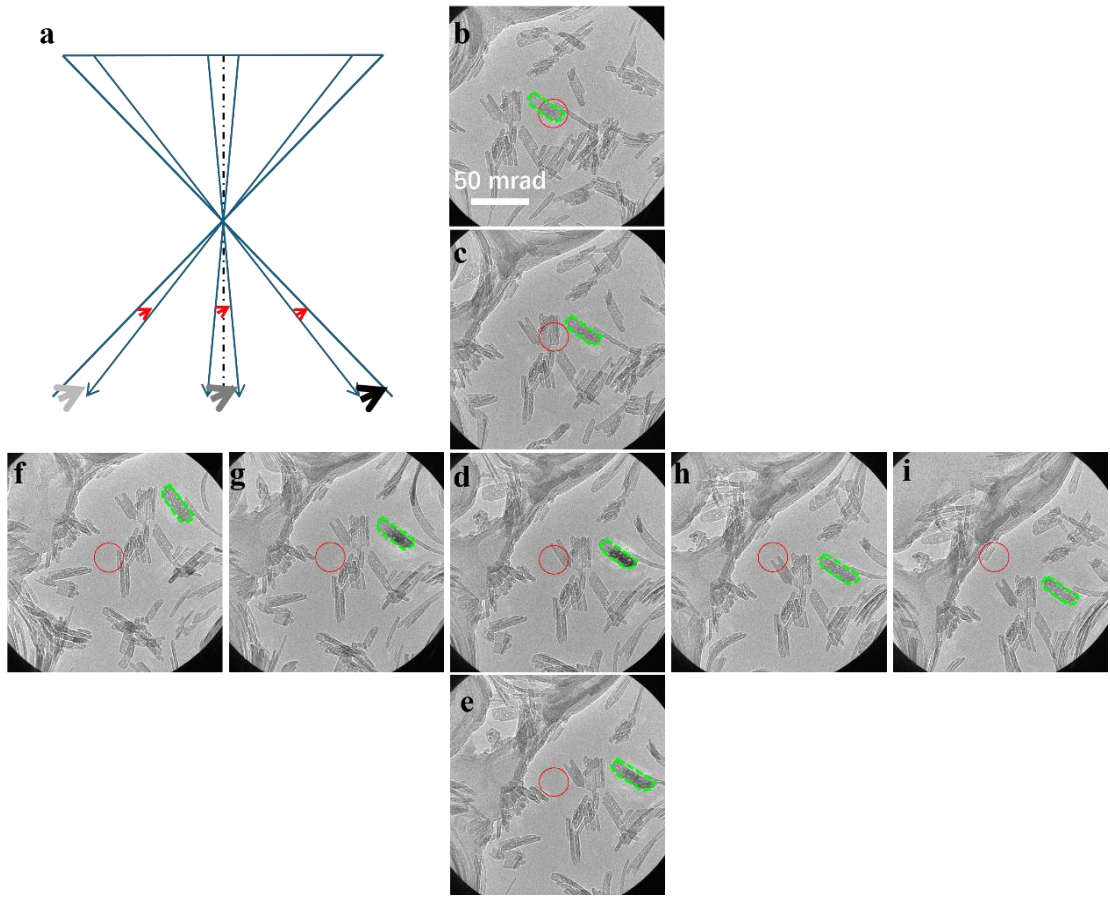

**Figure S2. Locating of the darkest shadow image position before tilting in Figure 2c.** (**a**) The electron ray diagram of the shadow image with an off-zone axis specimen. The darkest shadow image is not located on the optical center. (**b-i**) Experimental shadow images of the nanorod highlighted by green dotted lines. These images were acquired by in-plane moving the specimen holder. The contrast of the shadow images varies with different specimen positions. The darkest images appear in (d), which is same to Figure 2c in the main text.

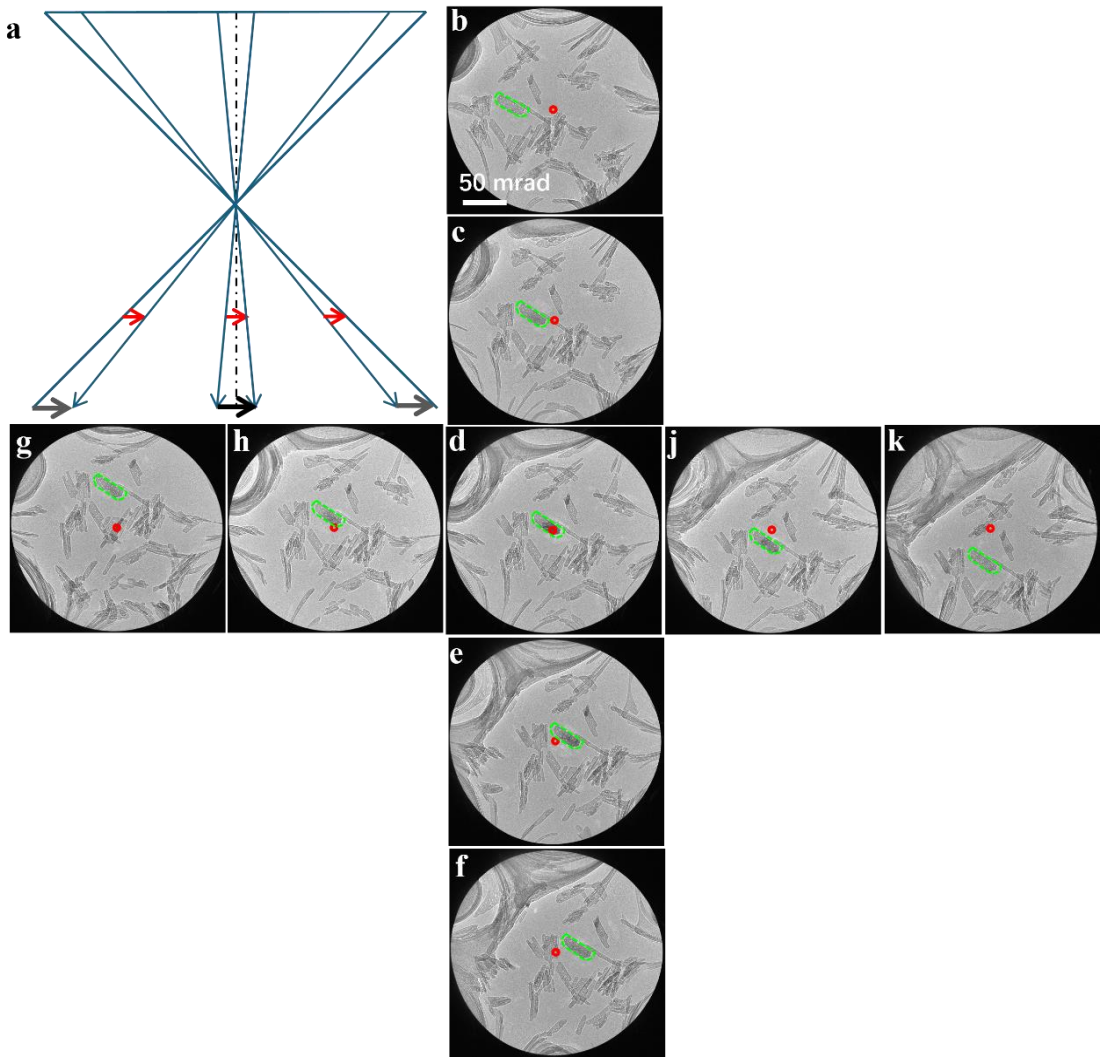

**Figure S3. Locating of the darkest shadow image position after tilting in Figure 2d.** (**a**) The electron ray diagram of the shadow image with a zone axis specimen. The darkest shadow image is located on the optical center. (**b-j**) Experimental shadow images of the nanorod highlighted by green dotted lines. These images were acquired by in-plane moving the specimen holder. The contrast of the shadow images varies with different specimen positions. The darkest image appears in (d), which is same to Figure 2c in the main text. After tiling, this darkest image is coincident to the optical center, indicating the specimen is tilted to the zone axis.

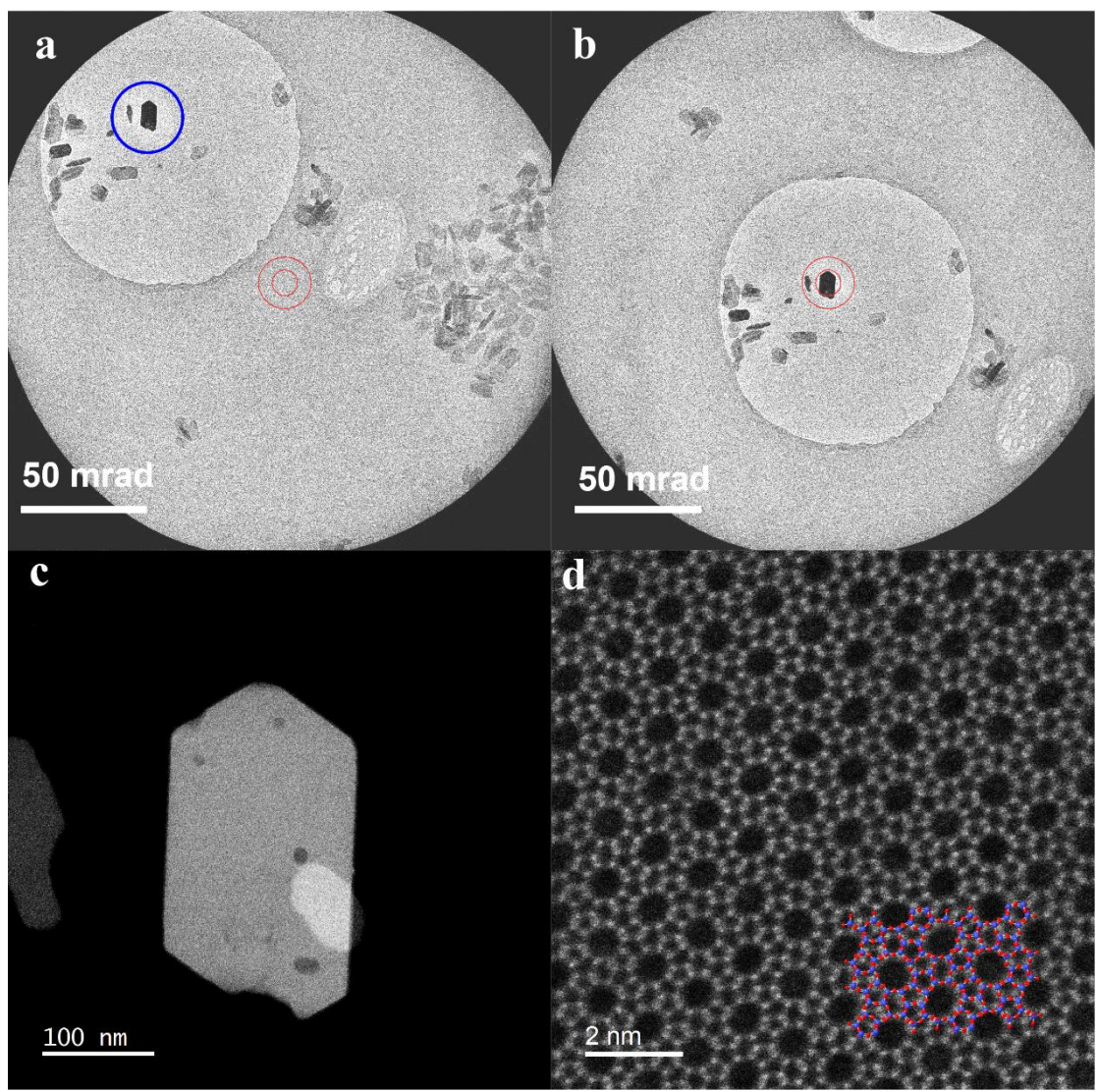

**Figure S4. Tilting of zeolite under low dose conditions.** (**a,b**) Experimental darkest shadow images of the ZSM-5 nanosheet in the center of the blue circle in (a) before (a) and after (b) tilting. (**c, d**) The ADF-STEM images of the nanosheet acquired after tilting. In (d), the atomic model of the ZSM-5 was overlaid on the image, where the blue and red spheres represent the Si and O atomic columns, respectively.

In this example, the probe current was set to ~115 pA, the defocus value was set to be ~30 μm. The estimated dose rate for tilting is ~0.17 e$^-$/Å$^2$/s.

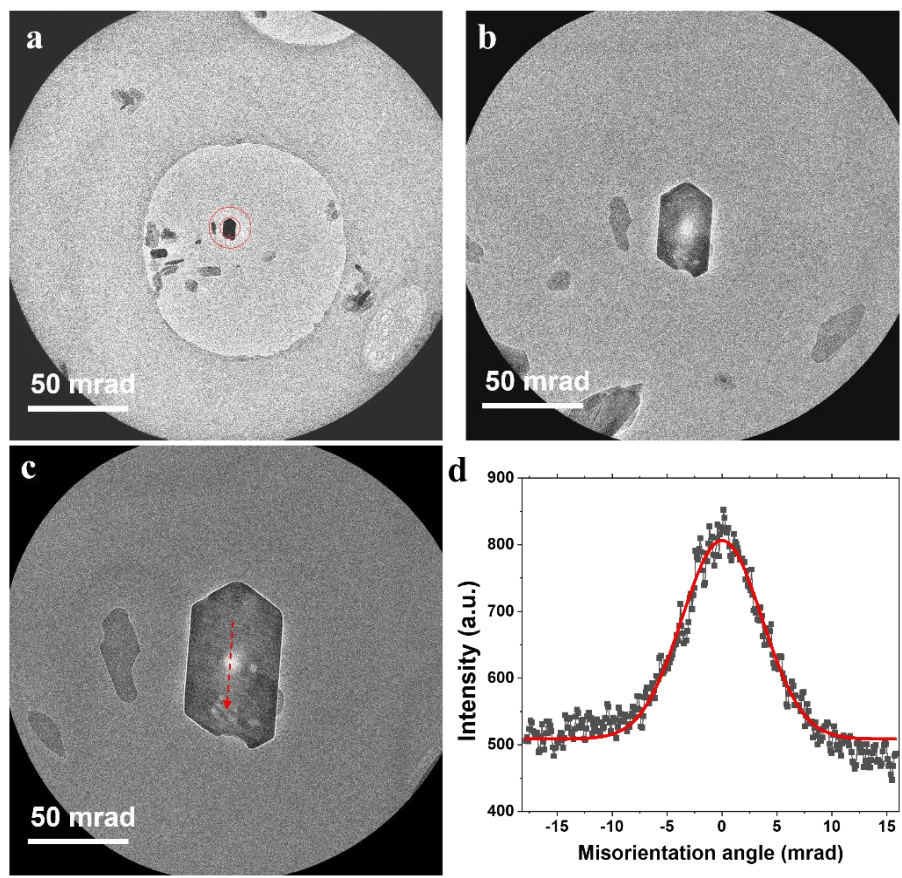

**Figure S5. Tilting of zeolite in Figure S4 with higher accuracy.** (**a**) The shadow image in Figure S4b. The defocus value was ~30 μm. (**b**) The shadow image acquired with the defocus of ~5.5 μm. (**c**) The shadow image acquired with the defocus of ~3 μm. (**d**) The intensity line profile along the red dotted line in (c). The line profile is fitted by a Gaussian function (the red curve). The standard error of the maximum intensity position is estimated to be 0.05 mrad.

It is important to note that the shadow images in Figure S5(b-c) show a brighter contrast at the zone position. This is because of the extinction distance[20], in which the diffracted beams would be scattered back to the transmitted beam, resulting a brighter contrast at the zone axis position of the shadow image. By fitting the intensity line profile, the standard error for determining the maximum intensity position is estimated to be ~0.05 mrad, which shows high accuracy to tilt the specimen to the zone axis.

## 2. Captions of Movies S1

**Movie S1. Real-time movie shows the automatically tilting.**

Automatically tilting the specimen to the zone axis involves the following steps:

1) Select the specimen to be tilt and move the specimen holder to find its darkest shadow image position. If you are using a convergence angle of 120 mrad and the specimen is less than ±6.9° off from the zone axis, its darkest shadow image will consistently appear within the field of view of the Ronchigram. If the darkest image of the specimen cannot be found in the Ronchigram, try to select an alternate sample.

2) Once the darkest image position is identified, press "Shift" and click the left mouse button at the position, triggering the script to automatically tilt the specimen to zone axis.